\documentclass{iopconfser}
\usepackage{cite}
\usepackage{amsmath,amssymb}
\usepackage{mathtools}
\usepackage{bm}
\usepackage{fix-cm}

\ifx\pdfoutput\undefined
\usepackage[dvipdfmx]{graphicx}
\usepackage[dvipdfmx]{hyperref}
\usepackage[dvipdfmx]{color}
\usepackage[dvipdfmx]{xcolor}
\else
\usepackage{graphicx}
\usepackage{hyperref}
\usepackage{color}
\usepackage{xcolor}
\fi

\newcommand{\bras}[1]{\langle#1\rvert}
\newcommand{\kets}[1]{\lvert#1\rangle}

\newcommand{\means}[1]{\langle#1\rangle}
\newcommand{\meansd}[1]{\langle\!\langle#1\rangle\!\rangle}

\hypersetup{colorlinks=true,linkcolor=blue,citecolor=blue,urlcolor=blue}

\begin{document}

\title{Impact of triplon damping on thermal Hall conductivity in Shastry-Sutherland model}

\author{Shinnosuke Koyama and Joji Nasu}

\affil{Department of Physics, Tohoku University, Sendai, Miyagi 980-8578, Japan}

\email{nasu@tohoku.ac.jp}

\begin{abstract}
We investigate the thermal Hall effect in the Shastry-Sutherland model, incorporating interactions between quasiparticle excitations.
In this model, with strong nearest-neighbor interactions, the ground state is well described by the direct product of spin-singlet states, and the elementary excitations to spin-triplet states are known as triplons.
In candidate materials for this model, Dzyaloshinskii-Moriya interactions are inevitably present, resulting in topologically non-trivial band structures for triplon excitations.
In this study, we examine quasiparticle damping due to triplon-triplon interactions as a potential factor contributing to the suppression of the thermal Hall effect.
We apply nonlinear flavor-wave theory to the Shastry-Sutherland model and treat triplons as bosonic excitations.
We calculate the triplon damping rate using the imaginary Dyson equation approach and evaluate the thermal Hall conductivity.
Our findings demonstrate that triplons with nonzero Berry curvature are scattered by thermally excited triplons.
This scattering effect suppresses the thermal Hall conductivity, particularly at finite temperatures, highlighting the significant role of triplon damping in the thermal Hall effect.
\end{abstract}

\section{Introduction}

Topological quantum states of matter have attracted considerable attention in condensed matter physics.
In particular, topologically nontrivial electronic band structures have been extensively studied in various systems, such as quantum Hall systems~\cite{thouless1982,kohmoto1985,hatsugai1993} and topological insulators~\cite{Kane2005,Fu2007,bernevig2006quantum}.
In these systems, topological properties manifest as quantum transport phenomena, such as the quantum Hall and spin Hall effects.
On the other hand, the topological properties of elementary excitations from symmetry-broken states have also been explored~\cite{sheng2006,zhang2010,zhang2011,katsura2010,matsumoto2011,qin2012,shindou2013,mook2014,matsumoto2014,li2016,zyuzin2016_prl,saito2019,wang2020,chen2021,ding2022,mcclarty2022,zhang2024thermal}.
Since these excitations are often described as charge-neutral bosonic quasiparticles, it is difficult to observe their topological properties electrically.
However, the thermal Hall effect, a thermal analog of the quantum Hall effect, has been proposed as a promising phenomenon to reflect the topological properties of bosonic quasiparticles~\cite{sheng2006,zhang2010,zhang2011,qin2012,katsura2010,matsumoto2011,mcclarty2022,zhang2024thermal}.
For example, thermal Hall effects originating from magnons and phonons have been observed in insulating magnets~\cite{onose2010,ideue2012,hirschberger2015_science,Akazawa2020,zhang2021_prl,hirschberger2015_prl,czajka2023}.
In these bosonic systems, the Berry curvature of the quasiparticle bands plays a crucial role in this effect~\cite{berry1984}.

In addition to magnons and phonons, which are collective modes arising from symmetry-broken phases, triplon excitations have been studied as another type of quasiparticle.
Triplons are spin-triplet excitations from a spin-singlet ground state in quantum magnets, often described as bosons~\cite{Sachdev1990}.
A typical example exhibiting triplon excitations is the spin-dimer phase of the Shastry-Sutherland model~\cite{shastry1981exact}, a two-dimensional quantum magnet with strong nearest-neighbor (NN) interactions.
As candidate materials for this system, SrCu$_2$(BO$_3$)$_2$ has been proposed~\cite{kageyama1999,Miyahara1999}, and its magnetic properties and excitation spectra have been actively studied~\cite{Koga2000,kodama2002magnetic,Corboz2013,Wang2018,shi2022discovery,nomura2023unveiling}.
In this compound, the magnetic properties cannot be fully explained by the pure Shastry-Sutherland model; additional interactions, such as the Dzyaloshinskii-Moriya (DM) interaction, are necessary to account for experimental results~\cite{Cepas2001,miyahara2004effects,Romhanyi2011}.
Due to the DM interaction, the triplon bands exhibit a topologically nontrivial band structure with nonzero Berry curvature, which induces the thermal Hall effect~\cite{romhanyi2015}.
However, recent experiments have shown no evidence of the thermal Hall effect in SrCu$_2$(BO$_3$)$_2$~\cite{suetsugu2022}.
As possible causes of this discrepancy, the effects of triplon-triplon interactions and coupling with phonons have been proposed, which were not considered in previous theoretical studies.
Recently, it has been suggested that interactions between bosonic quasiparticles play a crucial role in the stability of topological chiral edge modes~\cite{koyama2023,habel2024} and in the thermal Hall effect in magnon systems~\cite{koyama2024}, which may also apply to triplon systems.

In this paper, we focus on the finite-lifetime effect of triplons resulting from triplon-triplon interactions as a possible cause for the suppression of the thermal Hall effect.
We investigate the effect of triplon damping on the thermal Hall conductivity in the Shastry-Sutherland model incorporating DM interactions.
We introduce triplons as bosonic quasiparticles by applying the flavor-wave theory to this model.
We address contributions beyond the linear flavor-wave approximation by using the imaginary Dyson equation (iDE) approach.
To calculate the damping rate of triplons, we consider the collision and decay processes in the triplon self-energy.
Our findings reveal that at zero temperature, the higher-energy triplon bands are significantly broadened due to the damping effect, which originates from the decay process where a triplon splits into two quasiparticles.
At finite temperatures, the collision process involving thermally excited triplons also contributes to the damping effect.
Through this process, triplon bands with nonzero Berry curvature become smeared out.
We find that the thermal Hall conductivity is suppressed by triplon damping at finite temperatures, which originates from the scattering of triplons with nonzero Berry curvature by thermally excited triplons.
Furthermore, we demonstrate that these damping effects result from contributions beyond the lowest-order Born approximation, which can be addressed using the iDE approach.

This paper is organized as follows.
In Sec.~\ref{sec:model-method}, we describe the model and method employed in this study.
We introduce the Shastry-Sutherland model in Sec.~\ref{sec:model} and triplon excitations as its bosonic quasiparticles from the ground state based on the generalized Holstein-Primakoff transformation, which is referred to as the flavor-wave theory.
The linear approximation of the flavor-wave theory is explained in Sec.~\ref{sec:LFW}.
Contributions beyond the linear flavor-wave approximation are explored in Sec.~\ref{sec:NLFW}.
Nonlinear effects are addressed using the iDE approach and are incorporated by the triplon damping rate.
In Sec.~\ref{sec:thermal-Hall}, we present the method for calculating the thermal Hall conductivity with and without triplon damping.
Section~\ref{sec:result} discusses the results of this study.
In Sec.~\ref{sec:magnon-bands}, we show the triplon band structure and Berry curvature in the Shastry-Sutherland model within the linear flavor-wave approximation.
The results regarding the triplon damping effect on the spectral function are given in Sec.~\ref{sec:damping}.
The origin of the damping effect is examined through the scattering processes of triplons.
In Sec.~\ref{sec:THE-damping}, we present the thermal Hall conductivity calculated with triplon damping and discuss the impact of triplon damping at finite temperatures.
Finally, Sec.~\ref{sec:summary} is devoted to the summary.

\section{Model and method}
\label{sec:model-method}

\subsection{Shastry-Sutherland model}
\label{sec:model}

\begin{figure}[t]
    \begin{center}
    \includegraphics[width=\columnwidth,clip]{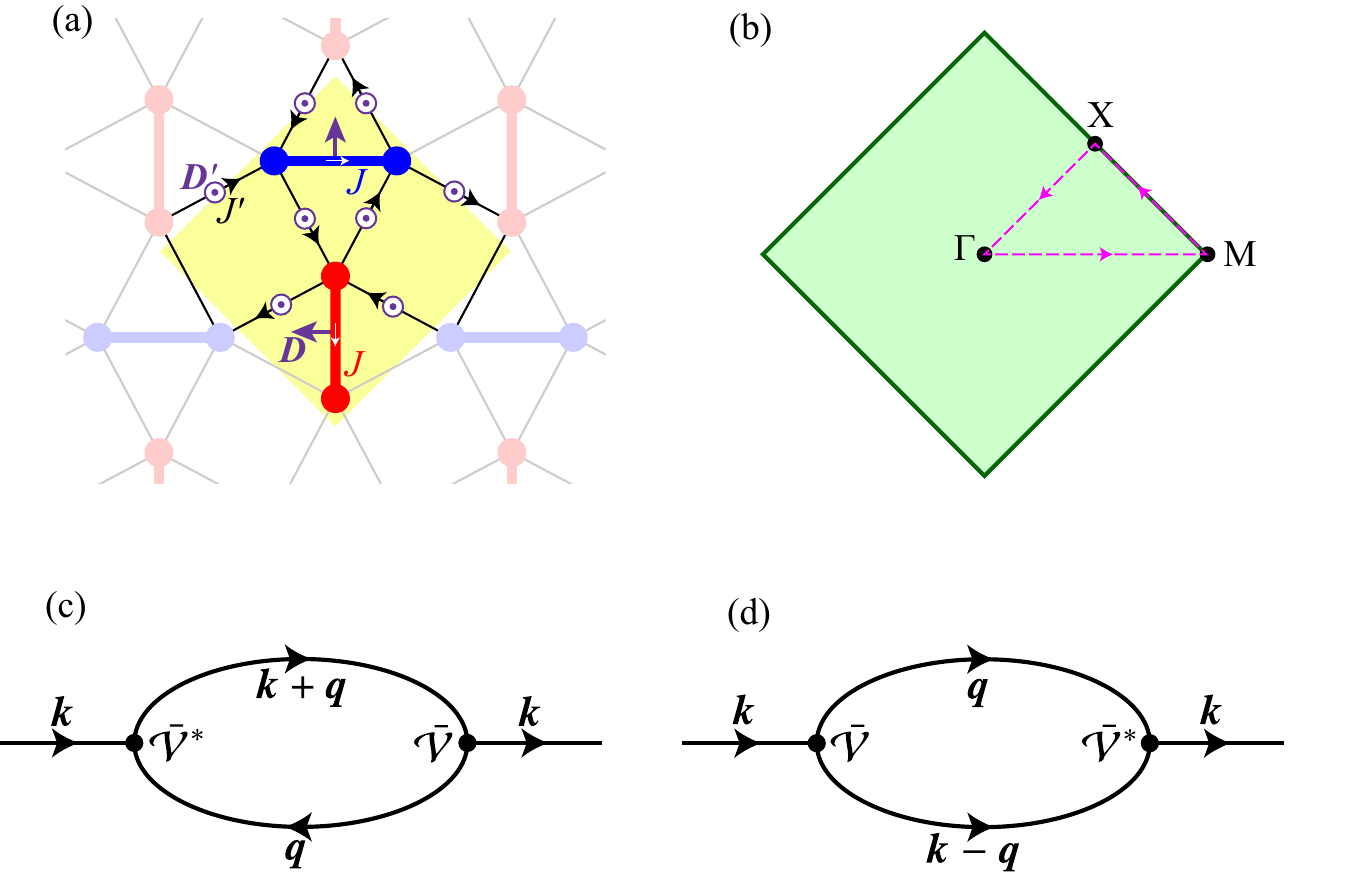}
    \caption{(a) Schematic picture of the Shastry-Sutherland lattice.
    The lattice consists of two types of dimers, depicted in red (dimer $A$) and blue (dimer $B$), with the antiferromagnetic Heisenberg interaction $J$ and DM interaction $\bm{D}$, where the bond directions are indicated by white arrows.
    These dimers are connected by bonds with Heisenberg-type exchange interaction $J'$ and DM interaction $\bm{D}'$.
    The bond directions for these connections are indicated by arrows similar to those for dimer bonds, represented with thick lines.
    The yellow area represents the unit cell of the lattice.
    (b) First Brillouin zone of the Shastry-Sutherland lattice.
    The dashed lines connect high-symmetry points within the Brillouin zone, which are used to plot the band structure of triplons in Figs.~\ref{fig:bc}, \ref{fig:spec}, \ref{fig:dos}, and \ref{fig:spec_diff}.
    (c),(d) Feynman diagrams in the self-energy of triplons for (c) the collision process and (d) the decay process.
    }
    \label{fig:lattice}
  \end{center}
\end{figure}

We consider the Shastry-Sutherland model, described by the following Hamiltonian:
\begin{align}
    {\cal H}=J\sum_{\means{i,i'}}\bm{S}_i\cdot\bm{S}_{i'}+\bm{D}\cdot\sum_{{\means{i,i'}}}\bm{S}_i\times\bm{S}_{i'}+J'\sum_{\meansd{i,i'}}\bm{S}_i\cdot\bm{S}_{i'}+\bm{D}'\cdot\sum_{\meansd{i,i'}}\bm{S}_i\times\bm{S}_{i'}-h\sum_i S_i^z,
\end{align}
where $\bm{S}_i$ is the spin-1/2 operator at site $i$, $J$ and $J'$ represent the antiferromagnetic Heisenberg exchange interactions between NN sites $\means{i,i'}$ and next-nearest-neighbor (NNN) sites $\meansd{i,i'}$, respectively. $\bm{D}$ and $\bm{D}'$ denote the DM interactions between NN and NNN sites, respectively, and $h$ is the magnetic field along the $z$ axis.
The DM vector $\bm{D}$ is oriented perpendicular to the NN bond direction and lies within the lattice plane.
On the other hand, $\bm{D}'$ can possess both in-plane and out-of-plane components~\cite{romhanyi2015}, but we neglect the in-plane component for simplicity.
The lattice structure of the Shastry-Sutherland model is depicted in Fig.~\ref{fig:lattice}(a).
In this figure, the directions of the DM interactions are indicated by purple arrows and circles, with the bond directions shown by white arrows for NN bonds and black arrows for NNN bonds.
In the unit cell, there are two types of dimers on NN bonds, labeled as $A$ and $B$ [see Fig.~\ref{fig:lattice}(a)].
The local Hamiltonian for dimer $j$ is expressed as
\begin{align}
    {\cal H}_j^{\rm dimer}=J\bm{S}_{j,1}\cdot\bm{S}_{j,2}+\bm{D}\cdot(\bm{S}_{j,1}\times\bm{S}_{j,2})-h(S_{j,1}^z+S_{j,2}^z),
\end{align}
where $\bm{S}_{j,1}$ and $\bm{S}_{j,2}$ are the spin operators for dimer $j$.
Then, the total Hamiltonian is rewritten as
\begin{align}\label{eq:hamiltonian}
    {\cal H}=\frac{1}{2}\sum_{j,j'}\sum_{\alpha,\alpha'}J_{jj'}^{\alpha\alpha'}S_{j}^\alpha S_{j'}^{\alpha'} + \sum_j{\cal H}_j^{\rm dimer},
\end{align}
where $\alpha$ and $\alpha'$ represent the composite indices $(\xi,\iota)$ for the spin component ($\xi=x,y,z$) and site index ($\iota=1,2$) within each dimer, and $J_{jj'}^{\alpha\alpha'}$ denotes the exchange interaction between dimers $j$ and $j'$, whose components are given by $J'$ and $\bm{D}'$.
Note that ${\cal H}_j^{\rm dimer}$ and $S_{j}^\alpha$ are represented as $4\times 4$ matrices in the subspace spanned by the direct product of the eigenstates for $S_{j,1}^z$ and $S_{j,2}^z$.

\subsection{Linear flavor-wave theory}
\label{sec:LFW}

We apply the mean-field (MF) approximation to the Hamiltonian given in Eq.~\eqref{eq:hamiltonian}.
The MF Hamiltonian for a dimer $j$ belonging to sublattice $A$ is expressed as~\cite{koyama2023}
\begin{align}
{\cal H}_j^{\rm MF}=\sum_{\alpha,\alpha'}\sum_{j'\in B}J_{jj'}^{\alpha\alpha'}S_{j}^\alpha \means{S^{\alpha'}}_B+{\cal H}_j^{\rm dimer},
\end{align}
where $\means{S^{\alpha'}}_B$ denotes the expectation value of $S^{\alpha'}$ on sublattice $B$.
The MF Hamiltonian for sublattice $B$ is introduced in a similar manner.
Since ${\cal H}_j^{\rm MF}$ is a $4\times 4$ matrix, it can be easily diagonalized to obtain the eigenvalues $E_l^m$ for $l=A,B$ and eigenstates $\kets{m;j}$ where $m=0,1,2,3$. Here, $\kets{0;j}$ represents the ground state of ${\cal H}_j^{\rm MF}$, and the others are the $m$-th excited states.
It should be noted that the eigenstate $\kets{m;j}$ is site-dependent; however, the eigenenergy and MF $\means{S^{\alpha'}}_l$ take the same values for all sites within the same sublattice.
Therefore, these values are distinguished by the sublattice index $l$.
The MFs $\means{S^{\alpha'}}_l=\bras{0;j}S^{\alpha'}\kets{0;j}$ for $j\in l$ are determined self-consistently.

Using the MF approximation, the original Hamiltonian is written as
\begin{align}
    {\cal H}=\sum_j {\cal H}_j^{\rm MF} + {\cal H}' + {\rm const.},
\end{align}
where ${\cal H}'$ represents the contribution beyond the MF approximation.
Here, we introduce the operators $X^{mm'}_{j}\equiv\kets{m;j}\bras{m';j}$ for dimer $j$.
The deviation from the MF Hamiltonian, ${\cal H}'$, is expressed by the off-diagonal components of $X^{mm'}_{j}$.
By employing the generalized Holstein-Primakoff transformation, we can express $X_{j}^{mm'}$ using bosonic operators, $a_{j,n}$ and $a_{j,n}^\dagger$, with $n=1,2,3$~\cite{joshi1999, kusunose2001,Lauchli2006,Tsunetsugu2006,Kim_flavor-wave2017, nasu2021, koyama2021,koyama2023}:
$X_{j}^{n0} = a_{j,n}^\dagger
\left(
  \mathcal{S} - \sum_{n'=1}^{3} a_{j,n'}^\dagger a_{j,n'}^{}
\right)^{1/2}$ and $X_j^{nn'} = a_{j,n}^\dagger a_{j,n'}^{}$, where $\mathcal{S}= X_j^{00}+ \sum_{n=1}^{3} a_{j,n}^\dagger a_{j,n}^{}$, which should be unity.
In the Shastry-Sutherland model, the three bosons introduced here are called triplons.
When the number of triplons is sufficiently small, the square root in $X_{j}^{n0}$ can be expanded with respect to $1/\mathcal{S}$.
Then, the Hamiltonian can also be expanded as
\begin{align}
    \label{eq:bosonic_H}
    \mathcal{H} =
      \mathcal{S}\left(\mathcal{H}_{0}+  \frac{1}{\sqrt{\mathcal{S}}}\mathcal{H}_3 + \frac{1}{\mathcal{S}}\mathcal{H}_4 + O(\mathcal{S}^{-3/2})
    \right) + \text{const.},
\end{align}
where $\mathcal{H}_{0}$ is a bilinear term of triplons, and $\mathcal{H}_3$ and $\mathcal{H}_4$ are three- and four-triplon terms, respectively.
When the higher-order terms beyond $\mathcal{H}_{0}$ are neglected, the Hamiltonian is reduced to that which describes free triplons.
This approach is referred to as the linear flavor-wave approximation.
By performing a Fourier transformation, we introduce $N$ bosons $a_{\bm{k},1}^{\dagger},a_{\bm{k},2}^{\dagger},\cdots , a_{\bm{k},N}^{\dagger}$ with $N=6$, as there are two dimers within the unit cell.
In the linear flavor-wave approximation, $\mathcal{H}_{0}$ is diagonalized via the Bogoliubov transformation as~\cite{colpa}
\begin{align}
    \label{eq:Hamil-B}
    \mathcal{H}_{0}= \frac{1}{2}\sum_{\bm{k}} \mathcal{B}^{\dagger}_{\bm{k}} \mathcal{E}_{\bm{k}} \mathcal{B}_{\bm{k}}^{},
\end{align}
where $\mathcal{B}_{\bm{k}}^{} = (b_{\bm{k},1},\cdots ,b_{\bm{k},N},b_{-\bm{k},1}^{\dagger} ,\cdots ,b_{-\bm{k},N}^{\dagger} )^T = T_{\bm{k}}^{-1} (a_{\bm{k},1},\cdots ,a_{\bm{k},N},a_{-\bm{k},1}^{\dagger} ,\cdots ,a_{-\bm{k},N}^{\dagger})^T$.
Note that $T_{\bm{k}}$ is a $2N\times 2N$ paraunitary matrix, satisfying $T_{\bm{k}}^{}\sigma_3 T_{\bm{k}}^\dagger = T_{\bm{k}}^{\dagger}\sigma_3 T_{\bm{k}}^{} = \sigma_3$
with the paraunit matrix
$
\sigma_3 =
\begin{psmallmatrix}
\bm{1}_{N\times N} & 0\\
0 & -\bm{1}_{N\times N}
\end{psmallmatrix}
$.
The matrix $\mathcal{E}_{\bm{k}}$ is diagonal and defined as
$\mathcal{E}_{\bm{k}} = \mathrm{diag}\{\varepsilon_{\bm{k},1}, \cdots, \varepsilon_{\bm{k},N}, \varepsilon_{-\bm{k},1}, \cdots, \varepsilon_{-\bm{k},N}\}$, where $\varepsilon_{\bm{k},\eta}$ with $\eta=1,\cdots, N$ represents the triplon energy, which must be positive.

\subsection{Nonlinear flavor-wave theory}
\label{sec:NLFW}

As the next step, we consider the effects of contributions beyond the linear flavor-wave approximation.
This study focuses on the damping effect on triplons, which arises from the imaginary part of the self-energy.
The lowest-order contribution to the self-energy is given by the second-order perturbation of the three-triplon Hamiltonian $\mathcal{H}_{3}$.
Among the interactions involving three triplons in $\mathcal{H}_{3}$, the following term contributes to the imaginary part of the self-energy:
\begin{align}
  \tilde{\mathcal{H}}_{3}=\frac{1}{2!}
  \sqrt{\frac{2}{N_{s}{\cal S}}} \sum_{\eta,\eta',\eta''}^N\sum_{\bm{k},\bm{q},\bm{p}}^{\bm{k}+\bm{q}=\bm{p}}
  \Bigl(
    \bar{\mathcal{V}}_{\bm{k},\bm{q}\leftarrow \bm{p}}^{\eta,\eta'\leftarrow\eta''} 
    b_{\bm{k},\eta}^\dagger b_{\bm{q},\eta'}^\dagger b_{\bm{p},\eta''}^{} + \text{H.c.}
  \Big),
\end{align}
where $N_s$ represents the number of dimers in the system, and $\bar{\mathcal{V}}_{\bm{k},\bm{q}\leftarrow \bm{p}}^{\eta,\eta'\leftarrow\eta''}$ is the bare vertex function~\cite{koyama2023}.
This term describes the triplon scattering process, where the triplon with momentum $\bm{k}$ and branch $\eta$ is scattered to momenta $\bm{q}$ and $\bm{p}$ with branches $\eta'$ and $\eta''$, respectively.
The summations over $\bm{k}$, $\bm{q}$, and $\bm{p}$ are taken in the first Brillouin zone under the restriction $\bm{p}=\bm{k}+\bm{q}$ due to momentum conservation, allowing for differences in reciprocal lattice vectors.

We evaluate the triplon self-energy $\Sigma_{\bm{k},\eta} (\omega)$ within the second-order perturbation of $\tilde{\mathcal{H}}_{3}$.
In this calculation, we focus solely on the imaginary part of the self-energy, which is related to triplon damping.
There are two contributions to the imaginary part of the self-energy: one is the collision process with thermally excited triplons, and the other is the decay process into two triplons, as diagrammatically illustrated in Figs.~\ref{fig:lattice}(c) and \ref{fig:lattice}(d), respectively.
It should be noted that the collision process contributes only at finite temperatures, whereas the decay process contributes even at zero temperature.
By considering these processes, we calculate the damping rate of triplons for the $\eta$-th branch $\Gamma_{\bm{k},\eta}$ using the iDE approach~\cite{chernyshev2009,maksimov2016_prb}.
In the lowest-order Born approximation, the damping rate is given by $-{\rm Im}\Sigma_{\bm{k},\eta} (\varepsilon_{\bm{k},\eta})$ using the self-energy  $\Sigma_{\bm{k},\eta} (\omega)$ originating from these processes.
In the iDE method, the damping rate is evaluated from $\Gamma_{\bm{k},\eta}=-\mathrm{Im}\Sigma_{\bm{k},\eta} (\omega^{*})$, where $\omega^{*}$ is the complex conjugate of the frequency $\omega$, introduced to satisfy causality~\cite{maksimov2016_prb}.
We solve the equation $\omega = \varepsilon_{\bm{k},\eta} + i\mathrm{Im} \Sigma_{\bm{k},\eta} (\omega^{*}, T)$ in a self-consistent manner.
It should be emphasized that this approach incorporates contributions beyond the lowest-order Born approximation, accounting for only one-loop diagrams shown in Figs.~\ref{fig:lattice}(c) and \ref{fig:lattice}(d) with the bare Green's function; in the iDE approach, the nonzero lifetime of triplons is included in the one-particle energy of the corresponding self-energy~\cite{chernyshev2009,maksimov2016_prb}.
This method has been successfully applied to various quantum spin systems, including frustrated quantum spin systems~\cite{chernyshev2009,maksimov2016_prb}, Kitaev-related models~\cite{winter2017_nc,koyama2023_NPSM}, and topological magnon systems with a chiral edge mode~\cite{koyama2024}.

\subsection{Thermal Hall conductivity}
\label{sec:thermal-Hall}

In this section, we introduce the method to evaluate the thermal Hall conductivity.
The thermal conductivity matrix $\kappa_{\lambda\lambda'}$ is defined in terms of the heat current $\bm{J}$ and the temperature gradient $\nabla T$ as
\begin{align}
    \label{eq:kxy-def}
    J_{\lambda} = \kappa_{\lambda\lambda'} (-\nabla_{\lambda'} T).
\end{align}
The thermal Hall conductivity $\kappa_{xy}^{H}$ is the antisymmetric component of the thermal conductivity matrix, defined as $\kappa_{xy}^{H} = (\kappa_{xy} - \kappa_{yx})/2$.
This quantity can be calculated using linear response theory.
In calculating thermal conductivity, it is necessary to incorporate contributions from heat magnetization in addition to those obtained from the Kubo formula~\cite{smrcka1977,cooper1997,xiao2006,qin2011}.
By considering these two contributions, the thermal Hall conductivity in a free triplon system described by ${\cal H}_0$ in Eq.~\eqref{eq:Hamil-B} is represented as~\cite{matsumoto2011,shindou2013,matsumoto2014,murakami2017}
\begin{align}
    \label{eq:kxy-free}
    \kappa_{xy}^{H;\mathrm{free}} 
    = -\frac{k_{\rm B}^2 T}{\hbar V} \sum_{\eta=1}^{N} \sum_{\bm{k}}\Omega_{\bm{k},\eta}
    c_2 (g(\varepsilon_{\bm{k},\eta})),
\end{align}
where $g(\varepsilon) = (e^{\beta \varepsilon} - 1)^{-1}$ represents the Bose distribution function with zero chemical potential, $c_{2}(x) = \int_{0}^{x} \left[ \ln (1+t)-\ln t \right]^2 dt$, and $\Omega_{\bm{k},\eta}$ is the Berry curvature of the bosonic bands, defined as
\begin{align}
    \Omega_{\bm{k},\eta}=i\left(\sigma_3 \frac{\partial T_{\bm{k}}^\dagger}{\partial k_x}\sigma_3 \frac{\partial T_{\bm{k}}}{\partial k_y}\right)_{\eta\eta} + {\rm c.c.}
\end{align}
In the presence of triplon damping, the thermal Hall conductivity is modified as follows~\cite{koyama2024}
\begin{align}
    \label{eq:kxy-L}
    \kappa_{xy}^{H} \simeq &-\frac{k_{\rm B}^2 T}{\hbar V}\sum_{\eta=1}^{N} \sum_{\bm{k}}\Omega_{\bm{k},\eta}
    \int_{-\infty}^{\infty}
    d\omega \rho_{\bm{k},\eta}(\omega)  c_2(g(\omega)),
\end{align}
where $\rho_{\bm{k},\eta}(\omega)$ represents the spectral function of triplons, defined by
\begin{align}\label{eq:spectral}
  \rho_{\bm{k},\eta}(\omega) = \frac{1}{\pi}\frac{\Gamma_{\bm{k},\eta}\theta(\omega)}{(\omega - \varepsilon_{\bm{k},\eta})^2 + \Gamma_{\bm{k},\eta}^2}.
\end{align}

For the following calculations, we set $J=1$ as the unit of energy and choose $J'=0.6J$, $|\bm{D}|=|\bm{D}'|=0.2J$, and $h=0.02J$.
In this study, to investigate the effects of triplon damping, the magnitudes of the DM interactions are set slightly larger than those in previous studies~\cite{Romhanyi2011,romhanyi2015,suetsugu2022}.

\section{Result}
\label{sec:result}

\subsection{Noninteracting magnon bands}
\label{sec:magnon-bands}

\begin{figure}[t]
  \begin{center}
  \includegraphics[width=0.73\columnwidth,clip]{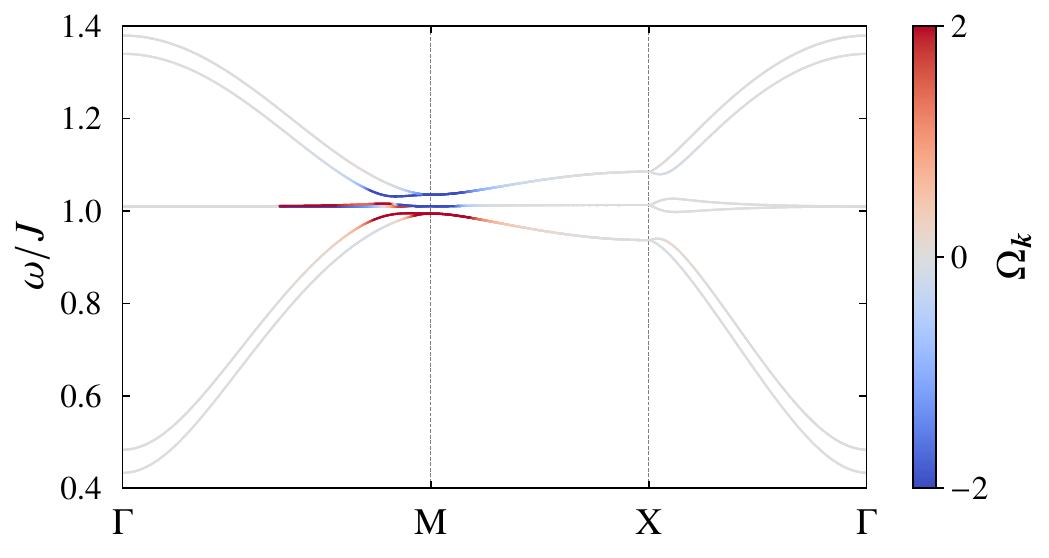}
  \caption{Berry curvature of the triplon bands, plotted along the dashed lines shown in Fig.~\ref{fig:lattice}(b).
  }
  \label{fig:bc}
\end{center}
\end{figure}

First, we examine the triplon band structure in the Shastry-Sutherland model without triplon damping.
Figure~\ref{fig:bc} shows the dispersion relation and Berry curvature of each triplon band.
There are six triplon branches, with $N=6$.
Two of these branches are nearly dispersionless and are located around $\omega\simeq J$, while the other four branches consist of dispersive bands.
Along the line from M to X, the triplon bands are doubly degenerate, with a nonzero Berry curvature.
This degeneracy is lifted along the lines from $\Gamma$ to M and from X to $\Gamma$, due to the DM interaction $D'$~\cite{Romhanyi2011}.

In Fig.~\ref{fig:bc}, the Berry curvature of the triplon bands is indicated by the line color.
As shown in this figure, the Berry curvature is nonzero, with a significant absolute value near the M point.
The lowest two branches exhibit positive Berry curvature, whereas the upper two branches exhibit negative Berry curvature.
The Berry curvature of the middle two nearly dispersionless branches changes its sign around the M point.
Despite the presence of double degeneracy, three sets of triplon bands are separated by energy gaps, indicating that the Chern number can be calculated for each set, which consists of two triplon branches.
We have verified that the Chern numbers for the three sets are $+2$, $0$, and $-2$, respectively, from the lower energy side, which is consistent with a previous study~\cite{romhanyi2015}.

\subsection{Effect of triplon damping}
\label{sec:damping}

\begin{figure}[t]
  \begin{center}
  \includegraphics[width=\columnwidth,clip]{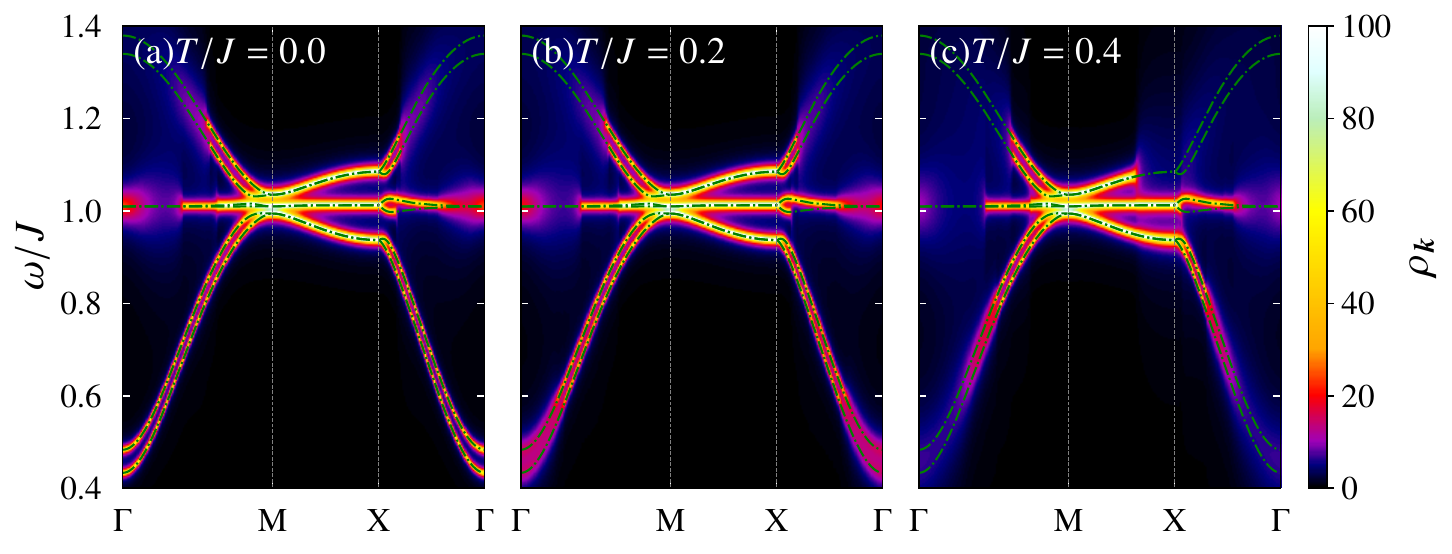}
  \caption{Spectral function of triplons, $\rho_{\bm{k}}(\omega)$ obtained by the iDE approach at (a) $T/J=0$, (b) $T/J=0.2$, and (c) $T/J=0.4$, plotted along the dashed lines shown in Fig.~\ref{fig:lattice}(b).
  }
  \label{fig:spec}
\end{center}
\end{figure}

In this section, we examine the effect of triplon damping.
Figure~\ref{fig:spec} shows the spectral function of triplons, $\rho_{\bm{k}}(\omega)$, as defined in Eq.~\eqref{eq:spectral} at several temperatures.
As shown in Fig.~\ref{fig:spec}(a), the high-energy part of the spectral function around the $\Gamma$ point is broadened due to triplon damping.
With an increase in temperature, the damping effect becomes more pronounced, and the spectral weight of the higher-energy branches weakens, even around the X point.
Furthermore, we observe that the low-energy part of the spectral function is smeared out by the damping effect only at finite temperatures.

\begin{figure}[t]
  \begin{center}
  \includegraphics[width=0.73\columnwidth,clip]{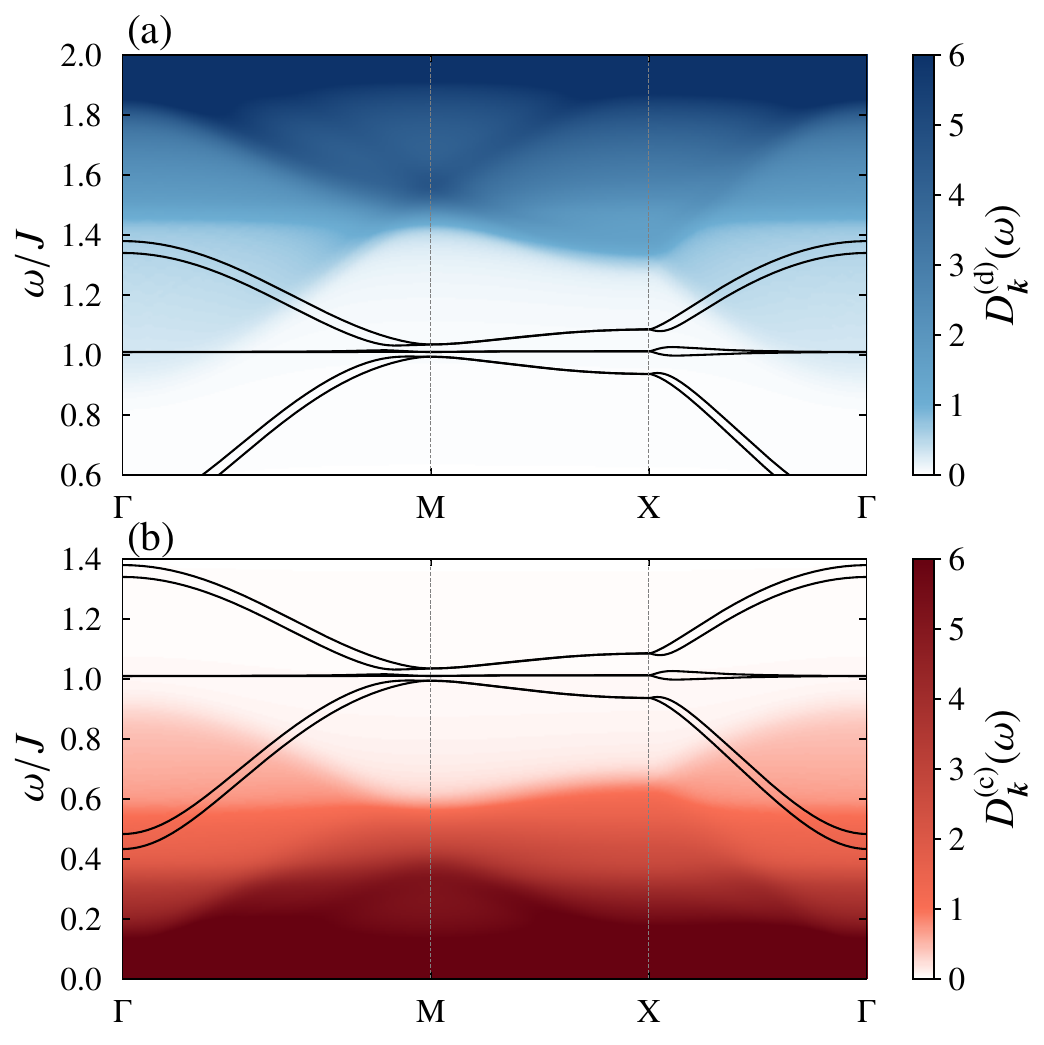}
  \caption{(a) Two-triplon DOS $D_{\bm{k}}^{\rm (d)}(\omega)$ and (b) two-triplon collision DOS $D_{\bm{k}}^{\rm (c)}(\omega)$, plotted along the dashed lines shown in Fig.~\ref{fig:lattice}(b).
  }
  \label{fig:dos}
\end{center}
\end{figure}

To understand this behavior, we introduce the two-triplon density of states (DOS) $D_{\bm{k}}^{\rm (d)}(\omega)$ and the two-triplon collision DOS $D_{\bm{k}}^{\rm (c)}(\omega)$, which are defined as
\begin{align}
  \label{eq:split_dos}
  D^{\text{(d)}}_{\bm{k}} (\omega) \equiv \frac{1}{N^2} \sum_{\eta',\eta''}^{N}\frac{2}{N_s}\sum_{\bm{q}} \delta(\omega -\varepsilon_{\bm{q},\eta'} - \varepsilon_{\bm{k}-\bm{q},\eta''}),
\end{align}
and
\begin{align}
  \label{eq:collision_dos}
  D^{\text{(c)}}_{\bm{k}} (\omega) \equiv  \frac{1}{N^2}\sum_{\eta',\eta''}^{N} \frac{2}{N_s}\sum_{\bm{q}}
  \delta(\omega + \varepsilon_{\bm{q},\eta'} - \varepsilon_{\bm{k}+\bm{q},\eta''}),
\end{align}
respectively.
Figure~\ref{fig:dos}(a) presents the two-triplon DOS $D_{\bm{k}}^{\rm (d)}(\omega)$.
When this DOS is nonzero at a triplon branch, the decay process of the triplon into two triplons, as depicted in Fig.~\ref{fig:lattice}(d), contributes to the damping of triplons under the lowest-order Born approximation.
As the original triplon bands appear above approximately $0.4J$, the two-triplon DOS becomes nonzero above twice this energy.
We observe that $D_{\bm{k}}^{\rm (d)}(\omega)$ overlaps with the higher-energy four-triplon bands near the $\Gamma$ point, resulting in a significant damping effect within the corresponding energy range at $T=0$, as illustrated in Fig.~\ref{fig:spec}(a).
At finite temperatures, the collision process of triplons with thermally excited triplons, shown in Fig.~\ref{fig:lattice}(c), is also allowed.
This effect can be understood from the overlap of $D_{\bm{k}}^{\rm (c)}(\omega)$ with the original triplon bands in Fig.~\ref{fig:spec}(b) within the lowest-order Born approximation.
As demonstrated in this figure, the bottom of the triplon bands around the $\Gamma$ point overlaps with $D_{\bm{k}}^{\rm (c)}(\omega)$, which leads to smearing of the lower-energy region of the spectral function at finite temperatures, as observed in Figs.~\ref{fig:spec}(b) and \ref{fig:spec}(c).

\subsection{Thermal Hall conductivity under triplon damping}
\label{sec:THE-damping}

\begin{figure}[t]
  \begin{center}
  \includegraphics[width=0.53\columnwidth,clip]{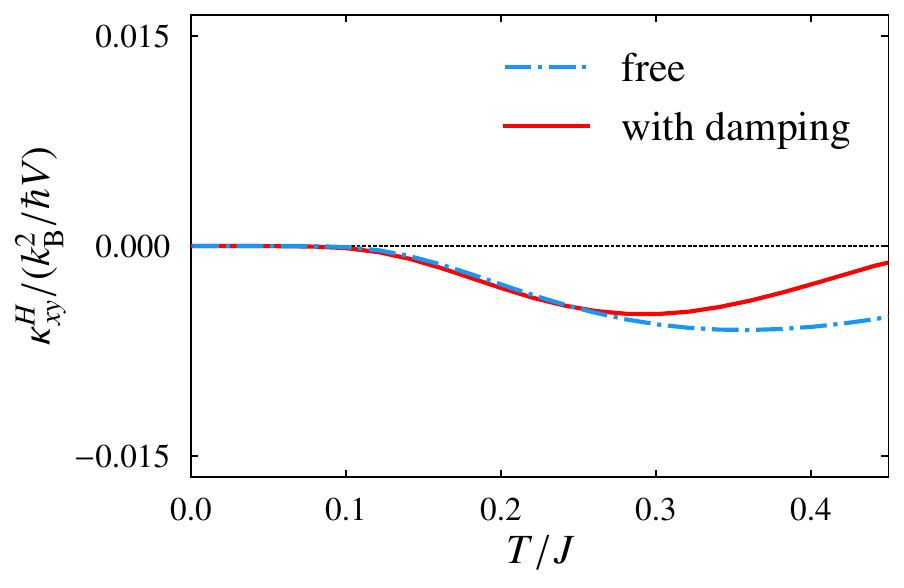}
  \caption{Temperature dependence of the thermal Hall conductivity $\kappa_{xy}^H$.
  The dashed-doted line represents the result obtained by the linear flavor-wave approximation, and the solid line is the result obtained by calculations including the triplon damping.
  }
  \label{fig:kxy}
\end{center}
\end{figure}

\begin{figure}[t]
  \begin{center}
  \includegraphics[width=0.7\columnwidth,clip]{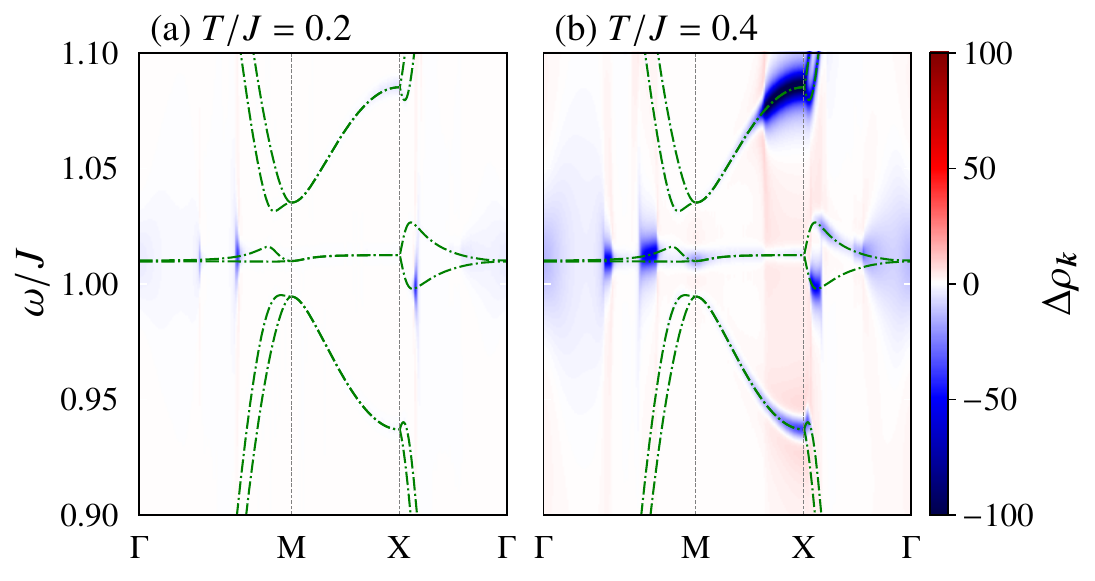}
  \caption{Difference in the spectral function $\Delta \rho_{\bm{k}}$ from the zero-temperature value at (a) $T/J=0.2$ and (b) $T/J=0.4$, plotted along the dashed lines shown in Fig.~\ref{fig:lattice}(b).
  }
  \label{fig:spec_diff}
\end{center}
\end{figure}

Finally, we examine the damping effect on the thermal Hall conductivity.
Figure~\ref{fig:kxy} shows the temperature dependence of the thermal Hall conductivity $\kappa_{xy}^H$ with and without triplon damping, calculated using Eqs.~\eqref{eq:kxy-L} and \eqref{eq:kxy-free}, respectively.
We find that the thermal Hall conductivity is significantly reduced by triplon damping, particularly above $T/J=0.3$.
From Eq.~\eqref{eq:kxy-L}, it is evident that the damping effect on triplons with nonzero Berry curvature plays a crucial role in suppressing the thermal Hall conductivity.
As shown in Fig.~\ref{fig:bc}, the Berry curvature of the triplon bands is prominent around the M point.
Thus, we focus on the spectral function around the M point to understand the damping effect on the thermal Hall conductivity.
Figure~\ref{fig:spec_diff} presents the difference in the spectral function $\Delta \rho_{\bm{k}}$ from the zero-temperature value, defined by $\Delta \rho_{\bm{k}}(\omega,T)=\rho_{\bm{k}}(\omega,T)-\rho_{\bm{k}}(\omega,T=0)$.
As shown in this figure, the suppression of the spectral weight on the nearly dispersionless bands becomes more pronounced around the M point at $T/J=0.4$ in comparison with $T/J=0.2$.
This suppression with increasing temperature is consistent with the reduction in the thermal Hall conductivity above $T/J=0.3$, as shown in Fig.~\ref{fig:kxy}.
It should be noted that there is no observable overlap between the magnon bands and $D_{\bm{k}}^{\rm (d)}(\omega)$ or $D_{\bm{k}}^{\rm (c)}(\omega)$ around the M point, as illustrated in Fig.~\ref{fig:dos}.
This result implies that contributions beyond the lowest-order Born approximation are necessary to account for the damping effect on the thermal Hall conductivity in the Shastry-Sutherland model.

\section{Summary}

\label{sec:summary}

In summary, we investigated the effect of triplon damping on the thermal Hall conductivity in the Shastry-Sutherland model.
By applying flavor-wave theory, we introduced triplons from the nearly spin-singlet ground state as bosonic excitations.
We also calculated the damping rate of triplons originating from scattering among triplons using the imaginary Dyson equation approach, which can incorporate contributions beyond the lowest-order Born approximation.
The damping effect significantly modulates the spectral function of triplons, particularly in the lower and higher-energy regions of the triplon bands.
Furthermore, we found that triplon damping also occurs around the M point, where the Berry curvature of the triplon bands is nonzero, at finite temperatures.
This effect leads to a suppression of the thermal Hall conductivity at finite temperatures, highlighting the impact of triplon damping on the thermal Hall effect.
On the other hand, recent experimental results reported that measured values are negligibly small compared to those obtained from the linear flavor-wave approximation.
This finding suggests that not only triplon damping but also other contributions, such as phonons, strongly influence the thermal Hall effect.
Incorporating the phonon contribution can be approached within our formalism, and thus such calculations remain for future work.

\section*{Acknowledgments}
The authors thank K.~Totsuka for fruitful discussions. 
Parts of the numerical calculations were performed in the supercomputing
systems in ISSP, the University of Tokyo.
This work was supported by Grant-in-Aid for Scientific Research from
JSPS, KAKENHI Grant No.~JP20H00122, JP22H01175, JP23H01129, JP23H04865, and JP24K00563.
and by JST, the establishment of university fellowships towards the creation of science technology innovation, Grant Number JPMJFS2102.

\bibliography{./refs}

\providecommand{\newblock}{}
\begin{thebibliography}{10}
\expandafter\ifx\csname url\endcsname\relax
  \def\url#1{{\tt #1}}\fi
\expandafter\ifx\csname urlprefix\endcsname\relax\def\urlprefix{URL }\fi
\providecommand{\href}[2]{#1}  
\providecommand{\eprint}[2][arXiv]{#1:\linebreak[0]#2}

\bibitem{thouless1982}
Thouless D~J, Kohmoto M, Nightingale M~P and den Nijs M 1982 {\em Phys. Rev. Lett.\/} {\bf 49} 405--408

\bibitem{kohmoto1985}
Kohmoto M 1985 {\em Ann. Phys.\/} {\bf 160} 343--354

\bibitem{hatsugai1993}
Hatsugai Y 1993 {\em Phys. Rev. Lett.\/} {\bf 71} 3697--3700

\bibitem{Kane2005}
Kane C~L and Mele E~J 2005 {\em Phys. Rev. Lett.\/} {\bf 95} 146802

\bibitem{Fu2007}
Fu L and Kane C~L 2007 {\em Phys. Rev. B\/} {\bf 76} 045302

\bibitem{bernevig2006quantum}
Bernevig B~A, Hughes T~L and Zhang S~C 2006 {\em Science\/} {\bf 314} 1757--1761

\bibitem{sheng2006}
Sheng L, Sheng D~N and Ting C~S 2006 {\em Phys. Rev. Lett.\/} {\bf 96} 155901

\bibitem{zhang2010}
Zhang L, Ren J, Wang J~S and Li B 2010 {\em Phys. Rev. Lett.\/} {\bf 105} 225901

\bibitem{zhang2011}
Zhang L, Ren J, Wang J~S and Li B 2011 {\em J. Phys.: Condens. Matter\/} {\bf 23} 305402

\bibitem{katsura2010}
Katsura H, Nagaosa N and Lee P~A 2010 {\em Phys. Rev. Lett.\/} {\bf 104} 066403

\bibitem{matsumoto2011}
Matsumoto R and Murakami S 2011 {\em Phys. Rev. Lett.\/} {\bf 106} 197202

\bibitem{qin2012}
Qin T, Zhou J and Shi J 2012 {\em Phys. Rev. B\/} {\bf 86} 104305

\bibitem{shindou2013}
Shindou R, Matsumoto R, Murakami S and Ohe J~i 2013 {\em Phys. Rev. B\/} {\bf 87} 174427

\bibitem{mook2014}
Mook A, Henk J and Mertig I 2014 {\em Phys. Rev. B\/} {\bf 90} 024412

\bibitem{matsumoto2014}
Matsumoto R, Sindou R and Murakami S 2014 {\em Phys. Rev. B\/} {\bf 89} 054420

\bibitem{li2016}
Li F~Y, Li Y~D, Kim Y~B, Balents L, Yu Y and Chen G 2016 {\em Nat. Commun.\/} {\bf 7} 12691

\bibitem{zyuzin2016_prl}
Zyuzin V~A and Kovalev A~A 2016 {\em Phys. Rev. Lett.\/} {\bf 117} 217203

\bibitem{saito2019}
Saito T, Misaki K, Ishizuka H and Nagaosa N 2019 {\em Phys. Rev. Lett.\/} {\bf 123} 255901

\bibitem{wang2020}
Li J, Wang L, Liu J, Li R, Zhang Z and Chen X~Q 2020 {\em Phys. Rev. B\/} {\bf 101} 081403

\bibitem{chen2021}
Chen Z~J, Wang R, Xia B~W, Zheng B~B, Jin Y~J, Zhao Y~J and Xu H 2021 {\em Phys. Rev. Lett.\/} {\bf 126} 185301

\bibitem{ding2022}
Ding Z~K, Zeng Y~J, Pan H, Luo N, Zeng J, Tang L~M and Chen K~Q 2022 {\em Phys. Rev. B\/} {\bf 106} L121401

\bibitem{mcclarty2022}
McClarty P~A 2022 {\em Annu. Rev. Condens. Matter Phys.\/} {\bf 13} 171--190

\bibitem{zhang2024thermal}
Zhang X~T, Gao Y~H and Chen G 2024 {\em Physics Reports\/} {\bf 1070} 1--59

\bibitem{onose2010}
Onose Y, Ideue T, Katsura H, Shiomi Y, Nagaosa N and Tokura Y 2010 {\em Science\/} {\bf 329} 297--299

\bibitem{ideue2012}
Ideue T, Onose Y, Katsura H, Shiomi Y, Ishiwata S, Nagaosa N and Tokura Y 2012 {\em Phys. Rev. B\/} {\bf 85} 134411

\bibitem{hirschberger2015_science}
Hirschberger M, Krizan J~W, Cava R~J and Ong N~P 2015 {\em Science\/} {\bf 348} 106--109

\bibitem{Akazawa2020}
Akazawa M, Shimozawa M, Kittaka S, Sakakibara T, Okuma R, Hiroi Z, Lee H~Y, Kawashima N, Han J~H and Yamashita M 2020 {\em Phys. Rev. X\/} {\bf 10} 041059

\bibitem{zhang2021_prl}
Zhang H, Xu C, Carnahan C, Sretenovic M, Suri N, Xiao D and Ke X 2021 {\em Phys. Rev. Lett.\/} {\bf 127} 247202

\bibitem{hirschberger2015_prl}
Hirschberger M, Chisnell R, Lee Y~S and Ong N~P 2015 {\em Phys. Rev. Lett.\/} {\bf 115} 106603

\bibitem{czajka2023}
Czajka P, Gao T, Hirschberger M, Lampen-Kelley P, Banerjee A, Quirk N, Mandrus D~G, Nagler S~E and Ong N~P 2023 {\em Nat. Mater.\/} {\bf 22} 36--41

\bibitem{berry1984}
Berry M~V 1984 {\em Proc. R. Soc. London. A. Math. Phys. Sci.\/} {\bf 392} 45--57

\bibitem{Sachdev1990}
Sachdev S and Bhatt R~N 1990 {\em Phys. Rev. B\/} {\bf 41} 9323--9329

\bibitem{shastry1981exact}
Shastry B~S and Sutherland B 1981 {\em Physica B+ C\/} {\bf 108} 1069--1070

\bibitem{kageyama1999}
Kageyama H, Yoshimura K, Stern R, Mushnikov N~V, Onizuka K, Kato M, Kosuge K, Slichter C~P, Goto T and Ueda Y 1999 {\em Phys. Rev. Lett.\/} {\bf 82} 3168--3171

\bibitem{Miyahara1999}
Miyahara S and Ueda K 1999 {\em Phys. Rev. Lett.\/} {\bf 82} 3701--3704

\bibitem{Koga2000}
Koga A and Kawakami N 2000 {\em Phys. Rev. Lett.\/} {\bf 84} 4461--4464

\bibitem{kodama2002magnetic}
Kodama K, Takigawa M, Horvatic M, Berthier C, Kageyama H, Ueda Y, Miyahara S, Becca F and Mila F 2002 {\em Science\/} {\bf 298} 395--399

\bibitem{Corboz2013}
Corboz P and Mila F 2013 {\em Phys. Rev. B\/} {\bf 87} 115144

\bibitem{Wang2018}
Wang Z and Batista C~D 2018 {\em Phys. Rev. Lett.\/} {\bf 120} 247201

\bibitem{shi2022discovery}
Shi Z, Dissanayake S, Corboz P, Steinhardt W, Graf D, Silevitch D, Dabkowska H~A, Rosenbaum T, Mila F and Haravifard S 2022 {\em Nat. Commun.\/} {\bf 13} 2301

\bibitem{nomura2023unveiling}
Nomura T, Corboz P, Miyata A, Zherlitsyn S, Ishii Y, Kohama Y, Matsuda Y, Ikeda A, Zhong C, Kageyama H {\em et~al\/} 2023 {\em Nat. Commun.\/} {\bf 14} 3769

\bibitem{Cepas2001}
C\'epas O, Kakurai K, Regnault L~P, Ziman T, Boucher J~P, Aso N, Nishi M, Kageyama H and Ueda Y 2001 {\em Phys. Rev. Lett.\/} {\bf 87} 167205

\bibitem{miyahara2004effects}
Miyahara S, Mila F, Kodama K, Takigawa M, Horvatic M, Berthier C, Kageyama H and Ueda Y 2004 {\em Journal of Physics: Condensed Matter\/} {\bf 16} S911

\bibitem{Romhanyi2011}
Romh\'anyi J, Totsuka K and Penc K 2011 {\em Phys. Rev. B\/} {\bf 83} 024413

\bibitem{romhanyi2015}
Romh{\'a}nyi J, Penc K and Ganesh R 2015 {\em Nat. Commun.\/} {\bf 6} 6805

\bibitem{suetsugu2022}
Suetsugu S, Yokoi T, Totsuka K, Ono T, Tanaka I, Kasahara S, Kasahara Y, Chengchao Z, Kageyama H and Matsuda Y 2022 {\em Phys. Rev. B\/} {\bf 105} 024415

\bibitem{koyama2023}
Koyama S and Nasu J 2023 {\em Phys. Rev. B\/} {\bf 108} 235162

\bibitem{habel2024}
Habel J, Mook A, Willsher J and Knolle J 2024 {\em Phys. Rev. B\/} {\bf 109} 024441

\bibitem{koyama2024}
Koyama S and Nasu J 2024 {\em Phys. Rev. B\/} {\bf 109} 174442

\bibitem{joshi1999}
Joshi A, Ma M, Mila F, Shi D~N and Zhang F~C 1999 {\em Phys. Rev. B\/} {\bf 60} 6584

\bibitem{kusunose2001}
Kusunose H and Kuramoto Y 2001 {\em J. Phys. Soc. Jpn.\/} {\bf 70} 3076--3083

\bibitem{Lauchli2006}
L\"auchli A, Mila F and Penc K 2006 {\em Phys. Rev. Lett.\/} {\bf 97} 087205

\bibitem{Tsunetsugu2006}
Tsunetsugu H and Arikawa M 2006 {\em J. Phys. Soc. Jpn.\/} {\bf 75} 083701

\bibitem{Kim_flavor-wave2017}
Kim F~H, Penc K, Nataf P and Mila F 2017 {\em Phys. Rev. B\/} {\bf 96} 205142

\bibitem{nasu2021}
Nasu J and Naka M 2021 {\em Phys. Rev. B\/} {\bf 103} L121104

\bibitem{koyama2021}
Koyama S and Nasu J 2021 {\em Phys. Rev. B\/} {\bf 104} 075121

\bibitem{colpa}
Colpa J 1978 {\em Physica\/} {\bf 93A} 327--353 ISSN 0378-4371

\bibitem{chernyshev2009}
Chernyshev A~L and Zhitomirsky M~E 2009 {\em Phys. Rev. B\/} {\bf 79} 144416

\bibitem{maksimov2016_prb}
Maksimov P~A and Chernyshev A~L 2016 {\em Phys. Rev. B\/} {\bf 93} 014418

\bibitem{winter2017_nc}
Winter S~M, Riedl K, Maksimov P~A, Chernyshev A~L, Honecker A and Valent{\'\i} R 2017 {\em Nat. Commun.\/} {\bf 8} 1152

\bibitem{koyama2023_NPSM}
Koyama S and Nasu J 2023 {\em New Physics: Sae Mulli\/} {\bf 73} 1123--1126

\bibitem{smrcka1977}
Smrcka L and Streda P 1977 {\em J. Phys. C\/} {\bf 10} 2153--2161

\bibitem{cooper1997}
Cooper N~R, Halperin B~I and Ruzin I~M 1997 {\em Phys. Rev. B\/} {\bf 55} 2344--2359

\bibitem{xiao2006}
Xiao D, Yao Y, Fang Z and Niu Q 2006 {\em Phys. Rev. Lett.\/} {\bf 97} 026603

\bibitem{qin2011}
Qin T, Niu Q and Shi J 2011 {\em Phys. Rev. Lett.\/} {\bf 107} 236601

\bibitem{murakami2017}
Murakami S and Okamoto A 2017 {\em J. Phys. Soc. Jpn.\/} {\bf 86} 011010

\end{thebibliography}

\end{document}